\documentclass[intlimits,twoside,a4paper]{article}
\usepackage{amsmath,amssymb}
\usepackage{graphicx}
\usepackage[T2A]{fontenc}
\usepackage[cp1251]{inputenc}

\usepackage[eqsecnum]{cmpj}

\issue{2011}{14}{1}{13003}

\doinumber{10.5488/CMP.14.13003}

\title[Inequalities in the AHM]{Generalized inequalities
for the Bogoliubov-Duhamel inner product with applications
in the Approximating Hamiltonian Method%
\thanks{The paper is dedicated to the 70th jubilee of N.N. Bogoliubov Jr.}}

\author[J.G.~Brankov, N.S.~Tonchev]{J.G.~Brankov\refaddr{label1,label2}
\thanks{E-mail: brankov@theor.jinr.ru}\,,
        N.S.~Tonchev\refaddr{label3}}

\authorcopyright{J.G. Brankov, N.S. Tonchev}

\addresses{
\addr{label1} Bogoliubov Laboratory of Theoretical Physics,
Joint Institute for Nuclear Research, \\141980 Dubna, Russian
Federation
\addr{label2} Institute of Mechanics, Bulgarian Academy of Science,
4 G. Bonchev Str., 1113 Sofia, Bulgaria
\addr{label3} Institute of Solid State Physics, Bulgarian Academy of Science,
72 Tzarigradsko Chaussee Blvd., \\1784 Sofia, Bulgaria
}

\date{Received November 29, 2010}

\begin{document}

 \maketitle

\begin{abstract}

Infinite sets of inequalities which generalize all the known
inequalities that can be used in the majorization step of the
Approximating Hamiltonian method are derived. They provide upper
bounds on the difference between the quadratic fluctuations of
intensive observables of a $N$-particle system and the
corresponding Bogoliubov-Duhamel inner product. The novel
feature is that, under sufficiently mild conditions, the upper
bounds have the same form and order of magnitude with respect to
$N$ for all the quantities derived by a finite number of
commutations of an original intensive observable with the
Hamiltonian. The results are illustrated on two types of exactly
solvable model systems: one with bounded separable attraction and
the other containing interaction of a boson field with matter.



\end{abstract}

\maketitle

\section{Introduction}

The Approximating Hamiltonian Method (AHM) provides a rigorous approach to the
study of some classes of statistical mechanical systems in the thermodynamic
limit. The method consists of the following interrelated ingredients:
\begin{itemize}
    \item[(i)] Description of classes of model systems which admit a rigorous
treatment in terms of a more simple approximating Hamiltonian;
  \item[(ii)] Rules according to which the approximating Hamiltonian is constructed
from the original one;
    \item[(iii)] Mathematical techniques for derivation of bounds which prove the
thermodynamic equivalence of the approximating and original Hamiltonians;
    \item[(iv)] Investigation of the thermodynamic and statistical properties of the
system described by the approximating Hamiltonian.
\end{itemize}

For the first time, the idea of the Approximating Hamiltonian
method (AHM) has been suggested on a heuristic level by N.N.
Bogoliubov in his paper on the theory of the weakly non-ideal
Bose-gas~\cite{Bog47}. He conjectured that, under the existence of
Bose condensate in the system, the normalized (by the square root
of the volume) creation/annihilation operators for bosons with
zero momentum can be replaced by $c$-numbers in the model
Hamiltonian. The value of these complex-conjugate numbers was
determined by thermodynamic arguments. Thus, the initial model
Hamiltonian was replaced by an approximating (trial) one, which
has the advantage to be easily diagonalized by a canonical $u$-$v$
transformation. The further development of the AHM took place in
the framework of the Bardeen-Cooper-Schrieffer (BCS) reduced model
of superconductivity~\cite{BCS}. In 1957 the corresponding
approximating Hamiltonian was suggested and diagonalized by means
of a $u$-$v$ canonical transformation~\cite{BZT57}. A little
later, N.N.~Bogoliubov~\cite{Bog60} rigorously proved that the
ground state energy density and the zero-temperature Green
functions for the model and approximating Hamiltonians coincide in
the thermodynamic limit.

The foundations of the modern formulation of the AHM have been
laid down by N.N.~Bogoliubov~(Jr.), see~\cite{Bog72} and
references therein. The essential generalization to classes of
quantum systems with separable interaction, considered at nonzero
temperatures, has been achieved by using the Bogoliubov
variational principle for the free energy density and a special
majorization technique based on integration over external sources.
Some major restrictions on the applicability of the method, such
as the quadratic form of the interaction Hamiltonian and its
boundedness, were removed by further extensions of the AHM, see
the review article~\cite{BBZKT84} and the books~\cite{BBZKT81,BDT}.

In this paper we generalize all the known inequalities which have
been, or could have been, used in the majorization step (iii) of
the AHM. They set different lower and upper bounds on the
quadratic fluctuations, proportional to the difference between the
original and approximating Hamiltonians, by terms involving new
functionals of which the Bogoliubov-Duhamel inner product is a
special case. Their place and role (not only in the AHM) is
illustrated by the derivation of some new consequences pertaining
to the infinitely coordinated anisotropic Heisenberg model and the
Dicke superradiance model. These inequalities can be used for
estimating the closeness of the Gibbs average values to the
corresponding Bogoliubov-Duhamel inner product for a special class
of observables. In some sense they are complementary to the
Bogoliubov inequality~\cite{Bog61} which has been used to exclude
conventional superfluid and superconducting long-range order in
one- and two-dimensional systems with gage invariant interactions.
By exploiting it for setting upper bounds on the spontaneous
magnetization or sublattice magnetization, Mermin and Wagner
\cite{MW} have rigorously proved the absence of long-range order
in one- and two-dimensional isotropic Heisenberg models with
finite-range interactions. Harris~\cite{Harris} has derived a
lower bound on the symmetrized average value of the product of an
operator $A$ and its conjugate $A^\dagger$, which is a special
case of the Bogoliubov inequality, but turns out to be sufficient
for the derivation of the result of Mermin and Wagner. Alternative
inequalities, setting a lower bound on the Bogoliubov-Duhamel
inner product, have been used in the proof of the existence of
spontaneous magnetization in a variety of quantum spin systems
\cite{Ro76,Ro77,DLS, S92}. In the physical literature, the
Bogoliubov-Duhamel inner product is also known as the
Bogoliubov-Kubo-Mori scalar product. In addition, it plays  an
important role in the linear response theory~\cite{NVW75}, Kondo
problem~\cite{Ro76}, the so-called parameter estimation problem in
quantum statistical mechanics and noncommutative probability
theory~\cite{PeTo93}, see also~\cite{Ru9} and references therein.

\section{Systems with bounded separable attraction}

Here we consider a special class of quantum statistical models
which are defined initially in a finite region $\Lambda$ of the
$d$-dimensional Euclidean space $R^d$ or the integer lattice
$Z^d$. By $|\Lambda|$ we denote the volume of $\Lambda$ in the
first case, or the number of lattice sites in the latter case. The
Hamiltonian ${\mathcal H}_{\Lambda}$ is defined as a self-adjoint
operator in a separable Hilbert space {\bf H}, and the
corresponding free energy density $f_{\Lambda}[{\mathcal
H}_{\Lambda}]$ is assumed to exist. For the sake of simplicity, we
do not explicitly distinguish between a Hamiltonian ${\mathcal
H}_{\Lambda}$, describing a system with fixed number of particles
$N$ in $\Lambda$, and the statistical operator ${\mathcal
H}_{\Lambda} - \mu {\mathcal N}$ in the grand canonical ensemble,
where $\mu$ is the chemical potential and ${\mathcal N}$ is the
number operator of particles. The density of the corresponding
thermodynamic potential is given by
\begin{equation}
f_{\Lambda}[\mathcal{H}_{\Lambda}] = - (\beta |\Lambda|)^{-1}\ln
Z[\mathcal{H}_{\Lambda}] \label{1c3},
\end{equation}
where $Z[\mathcal{H}_{\Lambda}]$ is the partition function; in all
cases the thermodynamic limit is denoted by $\mathrm{t-}\lim$. The
norm of a bounded operator $A$ is $\|A \|$, the symbol $A^{\#}$
remains for both the operator $A$ and its adjoint $A^{\dagger}$.
As usual, $[A,B]=AB-BA$ is the commutator of two operators.
Average values in the Gibbs ensemble with the Hamiltonian
${\mathcal H}_{\Lambda}$ are defined as
\begin{equation}
\langle \cdots \rangle_{\mathcal{H}_{\Lambda}} \equiv
\mathrm{Tr}({\mathrm e}^{-\beta \mathcal{H}_{\Lambda}}\cdots\,)
/Z[\mathcal{H}_{\Lambda}].
\end{equation}

Let the Hamiltonian of the system in $\Lambda$ be defined as
a sum of two self-adjoint operators,
\begin{equation}
{\mathcal H}_{\Lambda} = {\mathcal T}_{\Lambda} + {\mathcal
U}_{\Lambda}\,, \label{1Ham}
\end{equation}
where ${\mathcal T}_{\Lambda} = {\mathcal T}_{\Lambda}^{\dagger}$
is a trace-class operator which generates the Gibbs semigroup
$\{\exp(-\beta {\mathcal T}_{\Lambda})\}_{\beta \geqslant  0}$. In
addition, we impose the condition that the density of the
thermodynamic potential corresponding to ${\mathcal T}_{\Lambda}$
is bounded uniformly with respect to the volume of the system
$|\Lambda|$,
\begin{equation}
|f_{\Lambda}[{\mathcal T}_{\Lambda}]| \leqslant  M_0\,. \label{2.2}
\end{equation}

A distinguishing feature of the models with bounded separable
attraction is that the interaction Hamiltonian can be written as
an extensive self-adjoint operator of the form~\cite{Bog72}
\begin{equation}
{\mathcal U}_{\Lambda} = -|\Lambda|\sum_{s=1}^{n} g_{\mathrm{s}} A_{\Lambda,
s} A_{\Lambda, s}^{\dagger}\,.  \label{2.3}
\end{equation}
Here $g_{\mathrm{s}} > 0$, $s=1,\dots ,n$, are interaction
parameters and the intensive observables $A_{\Lambda,s}$,
$A_{\Lambda,s}^{\dagger}$ represent  uniformly bounded local
operators averaged over a region of the space (real space or
conjugate momentum space). The uniform boundedness of these
operators,
\begin{equation}
\|A_{\Lambda, s}\| = \|A_{\Lambda, s}^{\dagger}\|\leqslant  M_1\,,\qquad
s=1,\dots, n, \label{2.4}
\end{equation}
where the constant $M_1$ is independent of the volume $|\Lambda|$,
is essential for the applicability of the method in the case under
consideration. In the general framework one does not need explicit expressions
for the operators ${\mathcal T}_{\Lambda}$ and $A_{\Lambda, s}$.
It suffices to impose, in addition to (\ref{2.2}) and
(\ref{2.4}), the following general constraints ($s,s'=1,\dots, n$):
\begin{eqnarray}
\|[A_{\Lambda,s},{\mathcal T}_{\Lambda}] \| &\leqslant & M_{1,T}\,,\label{2.92} \\
\|[A_{\Lambda, s},A_{\Lambda, s'}^{\#}]\|&\leqslant & M_2 |\Lambda|^{-1}.
\label{2.93}
\end{eqnarray}

The heuristic rule for construction of the approximating
Hamiltonian ${\mathcal H}^{(0)}_{\Lambda}({\mathrm a})$ consists
in linearization of the original interaction Hamiltonian with
respect to the deviations of the intensive operators
$A_{\Lambda,s}$, $A_{\Lambda,s}^{\dagger}$ from some complex
numbers:
\begin{equation}
{\mathcal H}_{\Lambda}^{(0)}({\mathrm a})= {\mathcal
T}_{\Lambda}-|\Lambda | \sum_{s=1}^{n} g_{\mathrm{s}} (a_{\mathrm{s}}
A_{\Lambda,s}^{\dagger} +a_{\mathrm{s}}^{\star}A_{\Lambda,s} - a_{\mathrm{s}}
a_{\mathrm{s}}^{\star}), \label{2.21}
\end{equation}
where ${\mathrm a}=(a_1,\dots , a_n) \in {\mathbf C}^n$, and
$a_{\mathrm{s}}^{\star}$ is the complex conjugate of
$a_{\mathrm{s}}$. The latter numbers are considered to be
variational parameters, chosen so as to minimize the contribution
in the free energy density of the residual interaction Hamiltonian
\begin{equation}
{\mathcal H}_{\Lambda}^{(1)}({\mathrm a})\equiv {\mathcal
H}_{\Lambda}- {\mathcal H}_{\Lambda}^{(0)}({\mathrm a})=
-|\Lambda| \sum_{s=1}^{n} g_{\mathrm{s}} (A_{\Lambda,s}- a_{\mathrm{s}})
(A_{\Lambda,s}^{\dagger} -a_{\mathrm{s}}^{\star}) \leqslant  0. \label{2.22}
\end{equation}

The main result of the AHM for this class of model systems is
summarized in the absolute minimum principle for the approximating
free energy density:
\begin{equation}
0\leqslant  \min_{\mathrm a}f_{\Lambda}[{\mathcal H}_{\Lambda}^{(0)}({\mathrm a})]
      - f_{\Lambda}[{\mathcal H}_{\Lambda}] \leqslant
      \epsilon(|\Lambda|),
\label{epsattr}
\end{equation}
where $\epsilon(|\Lambda|) \rightarrow 0$ as $|\Lambda|\rightarrow
\infty$. This establishes the thermodynamic equivalence of the
free energy densities for the model Hamiltonian (\ref{1Ham}),
(\ref{2.3}), and the approximating one (\ref{2.21}).

The proof of (\ref{epsattr}) is carried out with the majorization
technique developed by Bogoliubov~Jr.~\cite{Bog72}. It starts with
the introduction of auxiliary external fields $\nu_{\mathrm{s}}$
and $\nu_{\mathrm{s}}^{\star}$ conjugate to the operators
$A_{\Lambda,s}^{\dagger}$ and $A_{\Lambda,s}$, respectively,
\begin{equation}
{\mathcal H}_{\Lambda}(\nu)\equiv {\mathcal H}_{\Lambda}-
|\Lambda| \sum_{s=1}^{n} (\nu_{\mathrm{s}} A_{\Lambda,s}^{\dagger} +
\nu_{\mathrm{s}}^{\star} A_{\Lambda,s}),  \qquad
{\mathcal H}_{\Lambda}^{(0)}({\mathrm a},\nu )\equiv
{\mathcal H}_{\Lambda}^{(0)}({\mathrm a})
-|\Lambda| \sum_{s=1}^{n} (\nu_{\mathrm{s}} A_{\Lambda,s}^{\dagger} +
\nu_{\mathrm{s}}^{\star} A_{\Lambda,s}).
\label{nu}
\end{equation}
Next, lower and upper bounds on the difference in the free energy
densities for the original and approximating Hamiltonians are set
by the Bogoliubov variational principle and the inequalities
following from it. In view of the non-positive definiteness of the
residual interaction Hamiltonian ${\mathcal
H}_{\Lambda}^{(1)}({\mathrm a})\equiv {\mathcal H}_{\Lambda}(\nu)
- {\mathcal H}_{\Lambda}^{(0)}({\mathrm a},\nu)$, the application
of the Bogoliubov inequalities  yields the bounds
\begin{eqnarray}
0& \leqslant&  \min_{\mathrm a} f_{\Lambda}[{\mathcal
H}_{\Lambda}^{(0)}({\mathrm a},\nu)]- f_{\Lambda}[{\mathcal
H}_{\Lambda}(\nu)] \nonumber \\%
 &\leqslant&   \sum_{s=1}^{n} g_{\mathrm{s}}
\langle(A_{\Lambda,s}- \langle A_{\Lambda,s}\rangle_{{\mathcal
H}_{\Lambda}(\nu)}) (A_{\Lambda,s}^{\dagger} -\langle
A_{\Lambda,s}^{\dagger}\rangle_{{\mathcal
H}_{\Lambda}(\nu)})\rangle_{{\mathcal H}_{\Lambda}(\nu)}\,,
\label{1c11}
\end{eqnarray}
valid for all complex fields $\nu_{\mathrm{s}}$, $s=1,2,\dots n$.

The fact that the absolute minimum of the free energy density
$f_{\Lambda} [{\mathcal H}_{\Lambda}^{(0)}({\mathrm a},\nu))]$ is
attained at a finite value $a= \bar{a}_{\Lambda}(\nu)$, which
depends on the thermodynamic parameters of the Gibbs ensemble as
well as on the size and shape of the domain $\Lambda$, has been
proved by using the conditions (\ref{2.2}) and (\ref{2.4}).

Here it is in place to mention that the attempt to prove directly
that the correlation functions in the right-hand side of
inequalities (\ref{1c11}) tends to zero as $|\Lambda| \rightarrow
\infty$ may happen to be an impossible task. An efficient means
for solving the problem provides the majorization technique of
Bogoliubov~(Jr.), see~\cite{Bog72}. Instead of direct evaluation
of the above correlation function, it uses inequalities which set
upper bounds to averages of the form
\begin{equation}
\langle \delta A_{\Lambda,s} \delta A_{\Lambda,s}^{\dagger}
\rangle_{{\mathcal H}_{\Lambda}(\nu)}\,, \qquad \delta A_{\Lambda,s}
\equiv A_{\Lambda,s} -\langle A_{\Lambda,s}\rangle_{{\mathcal
H}_{\Lambda}(\nu)}\,,
\label{kf}
\end{equation}
in terms involving second derivatives of the free energy density
with respect to external fields $\nu_{\mathrm{s}}$:
\begin{equation}
\frac{\partial^2 f_{\Lambda}[{\mathcal
H}_{\Lambda}(\nu)]}{\partial \nu_{\mathrm{s}}^\star \nu_{\mathrm{s}}} = -\beta |\Lambda|
(\delta A_{\Lambda,s};\delta A_{\Lambda,s}).
\label{second}
\end{equation}
Here $(A;B)$ denotes the Bogoliubov-Duhamel inner product on the
algebra of observables $A$, $B,\dots$, defined as
\begin{equation}
(A; B)_{\mathcal H} \equiv (Z [{\mathcal H}])^{-1}\int_0^{1}
{\mathrm d}\tau \: \mathrm{Tr} \left[{\mathrm e}^{-\beta(1
-\tau){\mathcal H}} A^{\dagger} {\mathrm e}^{-\beta \tau {\mathcal
H}} B\right] \label{1M}.
\end{equation}

In the remainder we consider the given observables $A, B, C,
\dots$ and fixed Hamiltonian ${\mathcal H}$ pertaining to a
quantum system in a finite region $\Lambda$. Whenever no confusion
could arise, for brevity of notation we will omit the subscripts
${\mathcal H}$ and $\Lambda$ and the argument of the partition
function $Z$.

\subsection{The Bogoluibov-Duhamel inner product}

The general properties of the functional (\ref{1M}) are considered
in the book~\cite{Ru69} and in the
articles~\cite{Ro76,Ro77,DLS,PeTo93,Ru9}. We warn the reader that
some authors use definitions which differ from~(\ref{1M}) by a
factor of $\beta$ and/or by involving the operator $A$ instead of
its adjoint $A^\dagger$. For our purposes it suffices to mention
the following. The inner product $(A;B)$ is conjugate symmetric,
$(A;B)= (B^\dagger ;A^\dagger)$, antilinear (linear) in the first
(second) argument, $(A+\alpha C;B)= (A;B) + \alpha^\star (C;B)$
($(A;B+\alpha C)= (A;B) + \alpha (A;C)$), it satisfies the
relationship
\begin{equation}
\beta (A; [{\mathcal H},B])_{\mathcal H}=\langle
[A^\dagger,B]\rangle_{\mathcal H}\,. \label{rel}
\end{equation}
The  inner product $(A;A)$ is nonnegative, $(A;A)\geqslant 0$, and
convex,
\begin{equation}
(A;A)\leqslant  (1/2)\langle AA^\dagger + A^\dagger A \rangle.
\label{conv}
\end{equation}
Finally, in the case when either $A$ or $B$ (or both) commute with
the Hamiltonian, the inner product $(A;B)$ reduces to the bilinear
complex-valued functional
\begin{equation}
\langle A ; B\rangle_{\mathcal H} =\langle
A^{\dagger}B\rangle_{\mathcal H}\,. \label{M5C}
\end{equation}

Note that in some works the operator metric in the algebra of
physical observables defined by~(\ref{M5C}) is called
Kubo-Martin-Schwinger metric, and the one defined by (\ref{1M}) is
called Bogoliubov-Kubo-Mori metric~\cite{Ru9,S2000}.

The further considerations are conveniently carried out by using
the spectral representation of the Bogoliubov-Duhamel inner
product. We assume that the Hamiltonian ${\mathcal H}$ has a
simple discrete spectrum only, $\{E_n, n=1,2,3\dots \}$ and denote
the corresponding eigenfunctions by $|n\rangle$, i.e., ${\mathcal
H}|n\rangle = E_n |n\rangle$, $n=1,2,3\dots$. By $A_{mn}=\langle
m|A|n\rangle$ we denote the corresponding matrix element of an
operator $A$. Then, the right-hand side of (\ref{1M}) can be
written as
\begin{eqnarray}
(A ;B)_{{\mathcal H}} = (Z_{\Lambda}
[{\mathcal H}])^{-1}\sum_{m,n}{}^\prime A^{*}_{mn}B_{mn}
\frac{\re^{-\beta E_{m}} - \re^{-\beta E_{n}}}
{\beta(E_{n}-E_{m})}%
+ (Z_{\Lambda}
[{\mathcal H}])^{-1}\sum_{n}{\mathrm e}^{-\beta E_{n}} A^{*}_{nn}B_{nn}\,,\quad
\label{M5}
\end{eqnarray}
where the prime in the double sum means that the term $n=m$ is
excluded.

Our aim is to majorize the quadratic fluctuations
\begin{equation}
\langle \delta A^{\dagger}\delta A\rangle =
\langle A^{\dagger} A\rangle_{\mathcal H} -|\langle A\rangle|^2
\label{Fluc3}
\end{equation}
by terms proportional to some power of the inner product
\begin{eqnarray}
(\delta A; \delta A)&=& (A;A)- |\langle A\rangle|^2 \nonumber \\ & =&
Z_{\Lambda}^{-1}
\sum_{m,n}{}^\prime |A_{mn}|^2\frac{{\mathrm e}^{-\beta E_{m}} -
{\mathrm e}^{-\beta E_{n}}}
{\beta(E_{n}-E_{m})}+ (Z_{\Lambda}[{\mathcal H}])^{-1}
\sum_{n}{\mathrm e}^{-\beta E_n} |A_{nn}|^2 -|\langle A\rangle|^2.\quad
\label{Fluc2}
\end{eqnarray}

Instead of (\ref{Fluc3}) it is more convenient to consider the symmetrized form
\begin{equation}
\frac{1}{2}\langle \delta A^{\dagger} \delta A + \delta A \delta A^{\dagger}\rangle
 = \frac{1}{2}\langle A^{\dagger} A +  A A^{\dagger} \rangle -|\langle A\rangle|^2  .
\label{FlucS}
\end{equation}
Here we have
\begin{eqnarray}
\frac{1}{2}\langle A^{\dagger} A +  A A^{\dagger}\rangle &=& Z^{-1}
\sum_{m,n}{\mathrm e}^{-\beta E_{n}}\frac{1}{2}(|A_{mn}|^2 + |A_{nm}|^2) \nonumber \\
& = &Z^{-1}\sum_{m,n}{}^\prime |A_{mn}|^2\frac{1}{2}({\mathrm e}^{-\beta E_{n}}
+ {\mathrm e}^{-\beta E_{m}})+
Z^{-1}\sum_{n} |A_{nn}|^2{\mathrm e}^{-\beta E_{n}}.
\label{Fluc4}
\end{eqnarray}

By comparing equations~(\ref{Fluc2}), (\ref{FlucS}) and (\ref{Fluc4}) we obtain
\begin{eqnarray}
\frac{1}{2}\langle A^{\dagger} A + A A^{\dagger}\rangle - (A; A)
= Z^{-1}
\sum_{m,n}{}^\prime |A_{mn}|^2\left\{\frac{1}{2}({\mathrm e}^{-\beta E_{n}}
+ {\mathrm e}^{-\beta E_{m}})
- \frac{{\mathrm e}^{-\beta E_{m}} - {\mathrm e}^{-\beta E_{n}}}
{\beta(E_{n}-E_{m})}\right\}.
\label{dif}
\end{eqnarray}

Now, by using the identity
\begin{equation}
{\mathrm e}^{-\beta E_{m}} + {\mathrm e}^{-\beta E_{n}}=
({\mathrm e}^{-\beta E_{n}}-{\mathrm e}^{-\beta E_{m}})\coth
\frac{\beta(E_{m}-E_{n})}{2}
\label{coth1}
\end{equation}
we can express the difference (\ref{dif}) in two equivalent forms:
\begin{equation}
\frac{1}{2}\langle A^{\dagger} A + A A^{\dagger}\rangle - (A; A)
 = Z^{-1}
\sum_{m,n}{}^\prime |A_{mn}|^2\frac{{\mathrm e}^{-\beta E_{m}} -
{\mathrm e}^{-\beta E_{n}}}
{\beta(E_{n}-E_{m})}(X_{mn}\coth X_{mn} -1),
\label{dif1}
\end{equation}
and
\begin{equation}
\frac{1}{2}\langle A^{\dagger} A + A A^{\dagger}\rangle - (A; A)
= Z^{-1}\sum_{m,n}{}^\prime |A_{mn}|^2\frac{1}{2}({\mathrm e}^{-\beta E_{n}} +
{\mathrm e}^{-\beta E_{m}}) \left(1-\frac{1}{X_{mn}\coth X_{mn}}\right),
\label{dif2}
\end{equation}
where
\begin{equation}
X_{mn}= \frac{1}{2}\beta (E_{m}-E_{n})  .
\label{X}
\end{equation}

The application of the elementary inequality $x\coth x \geqslant 1$ to
the right-hand side of (\ref{dif1}) immediately yields the
convexity property (\ref{conv}).

Different choices of the  upper bound on the right-hand side of
(\ref{dif1}) generate different inequalities. Thus, the
 inequality of Brooks Harris~\cite{Harris},
\begin{equation}
(A;A)\leqslant  \frac{1}{2}\langle AA^+ + A^+A\rangle \leqslant
(A;A)+\frac{\beta}{12}\langle [[A^+,{\mathcal H}],A]\rangle
\label{Harris}
\end{equation}
 is obtained by setting
\begin{equation}
1\leqslant  x\coth x \leqslant  1+\frac{1}{3}x^2 \label{elemBH}.
\end{equation}

On the other hand, if one uses another elementary inequality,
\begin{equation}
1\leqslant  x\coth x \leqslant  1+|x|, \label{elemG}
\end{equation}
and subsequently  applies the H\"{o}lder inequality, one obtains
the result due to Ginibre~\cite{G}:
\begin{equation}
(A;A)\leqslant  \frac{1}{2}\langle AA^{+} + A^{+}A\rangle\leqslant  (A;A)+
\frac{1}{2}\{(A;A)\beta \langle [[A^+,{\mathcal
H}],A]\rangle\}^{\frac{1}{2}} \label{G}.
\end{equation}

A different choice of the parameters in the H\"{o}lder inequality,
followed by the implementation of the upper bound
\begin{equation}
|\re^{-\beta E_{l}} - \re^{-\beta E_{m}}| < |\re^{-\beta E_{l}} +
\re^{-\beta E_{m}}|, \label{ub}
\end{equation}
generates a symmetric version of the inequality due to Bogoliubov
(Jr.)~\cite{Bog72}:
\begin{equation}
\frac{1}{2}\langle AA^+ + A^+A\rangle \leqslant
(A;A)+\frac{1}{2}[(A;A)\beta]^{2/3}\{\langle [A^+,{\mathcal
H}][{\mathcal H}, A]+ [{\mathcal H}, A][A^+,{\mathcal
H}]\rangle\}^{1/3}. \label{Jr}
\end{equation}

Due to relation (\ref{second}), each of the above inequalities can
be used in the AHM to majorize the quadratic fluctuations
(\ref{kf}) in terms involving second derivatives of the free
energy density with respect to the external fields $\nu_{\mathrm{s}}$.
However, note that in the zero temperature limit $\beta
\rightarrow \infty$ the right-hand side of the simplest inequality
(\ref{Harris}) diverges.

\section{Main inequalities}

To derive generalizations of the known inequalities involving the
Bogoliubov-Duhamel  inner product, a set of new functionals
$F_{k}(J;J)$, $k=0,1,2,\dots$, is defined by their spectral
representation:
\begin{equation}
F_{k}(J;J):=Z^{-1} \sum_{ml}|J_{ml}|^{2}|\re^{-\beta E_{l}} -
(-1)^{k}\re^{-\beta E_{m}}|(\beta|E_{m}-E_{l}|)^{k-1}.
\label{BT2010}
\end{equation}
The specific choice of $k=0,1,2,\dots $ is motivated by the relation
of functionals (\ref {BT2010}) to the Gibbs average values of some
commutators and anticommutators involving the  given operators $J$
and Hamiltonian ${\mathcal H}$. Indeed,
\begin{itemize}
\item[(i)] If $k=2n$, $n=0,1,2,3,\dots$, then
\begin{eqnarray}
F_{2n}(J;J) &\equiv& Z^{-1} \sum_{ml}|J_{ml}|^{2}|\re^{-\beta E_{l}}
- \re^{-\beta E_{m}}|(\beta|E_{m}-E_{l}|)^{2n-1}
\nonumber\\&=&\beta^{2n}(R_{n};R_{n})=
\beta^{2n-1}\langle[R_{n}^{+}R_{n-1}-R_{n-1}R_{n}^{+}]\rangle,
\label{BT10}
\end{eqnarray}
where, by definition, $R_{-1}\equiv X_{J{\mathcal H}}$ is a
solution of the operator equation $J = [X_{J{\mathcal
H}},{\mathcal H}]$, and
\begin{equation}
R_{0}\equiv R_{0}(J)= J,\quad  R_{1}\equiv R_{1}(J) =[{\mathcal
H},J], \quad R_{n} \equiv R_{n}(J)= [{\mathcal H},R_{n-1}(J)],
\quad n=1,2,3, \dots \label{R}
\end{equation}
These observables have been introduced in~\cite{BPR}.
\item[(ii)]
If $k=2n+1$, $n=0,1,2,3,\dots$, then
\begin{eqnarray}
F_{2n+1}(J;J)&\equiv&  Z^{-1}\sum_{ml}|J_{ml}|^{2}(\re^{-\beta
E_{l}} + \re^{-\beta E_{m}})[\beta(E_{m}-E_{l})]^{2n}
\nonumber\\&=&\beta^{2n}\langle[R_{n}R_{n}^{+}+R_{n}^{+}R_{n}]\rangle.
\label{M911}
\end{eqnarray}
\end{itemize}

In particular,
\begin{eqnarray}
&&F_{0}(J;J)= (J;J), \qquad F_{1}(J;J)= \langle JJ^+ + J^+J\rangle,
\qquad F_{2}(J;J)= \beta \langle [[J^+,{\mathcal H}],J]\rangle,
\nonumber \\&& F_{3}(J;J)= \beta^2 \langle [J^+,{\mathcal
H}][{\mathcal H}, J]+ [{\mathcal H}, J][J^+,{\mathcal H}]\rangle.
\label{Part}
\end{eqnarray}

The functionals (\ref{BT2010}) will be used to generalize all the
known inequalities used in the AHM.

\subsection{Generalization of the Harris inequality}

By using the identity (\ref{coth1}),
one can rewrite the equality (\ref{M911}) in the form
\begin{equation}
F_{2n+1}(J;J)\equiv  Z^{-1}\sum_{ml}|J_{ml}|^{2}(\re^{-\beta E_{l}}
- \re^{-\beta E_{m}})\coth
\frac{\beta(E_{m}-E_{l})}{2}[\beta(E_{m}-E_{l})]^{2n} .
\label{M911M}
\end{equation}
Now, from the elementary inequalities (\ref{elemBH})
it follows that
\begin{equation}
F_{2n}(J;J)\leqslant  \frac{1}{2}F_{2n+1}(J;J)\leqslant
F_{2n}(J;J)+\frac{1}{12}F_{2n+2}(J;J) . \label{BHgen}
\end{equation}

This is a generalization of the Brooks Harris inequality
(\ref{Harris}), since the latter is recovered in the particular
case of $n=0$.

\subsection{Generalization of the Plechko inequalities}

The application of the elementary inequalities (\ref{elemG})
to the right-hand side of (\ref{M911M}) leads to
\begin{equation}
F_{2n}(J;J)\leqslant  \frac{1}{2}F_{2n+1}(J;J)\leqslant  F_{2n}(J;J)+
(2Z)^{-1}\sum_{ml}|J_{ml}|^{2}|\re^{-\beta E_{l}} - \re^{-\beta
E_{m}}|[\beta(E_{m}-E_{l})]^{2n}. \label{genG}
\end{equation}

The problem now is that, due to the absolute value of the
difference of the two Gibbs exponents, there is no apparent
interpretation of the sum in the right-hand side of~(\ref{genG})
in terms of average values. A known way to overcome this
difficulty is based on the application of the H\"{o}lder
inequality
\begin{equation}
\sum_{k}|x_{k}y_{k}| \leqslant
\left(\sum_{k}|x_{k}|^{p}\right)^{1/p}\left(\sum_{k}
|y_{k}|^{q}\right)^{1/q}, \qquad p,q > 1,\qquad  1/p + 1/q = 1.
\label{Holder}
\end{equation}
By setting $k = (m,l)$ and
\begin{eqnarray}
&&|x_{ml}| = \left\{|J_{ml}|^2 \frac {\re^{-\beta E_{l}} - \re^{-\beta
E_{m}}}{\beta(E_{m}-E_{l})}\right\}^{1/p}, \nonumber \\
&&|y_{ml}| = \left\{|J_{ml}|^2 \frac {\re^{-\beta E_{l}} - \re^{-\beta
E_{m}}}{\beta(E_{m}-E_{l})}[\beta|E_{m}-E_{l})|]^{(2n+1)q}\right\}^{1/q},
\label{H1}
\end{eqnarray}
we obtain
\begin{eqnarray}
\lefteqn{(2Z)^{-1}\sum_{ml}|J_{ml}|^{2}|\re^{-\beta E_{l}} - \re^{-\beta
E_{m}}|[\beta(E_{m}-E_{l})]^{2n}}\nonumber \\ &&\leqslant
\frac{1}{2}(J;J)^{1/p}\left\{Z^{-1}\sum_{ml}|J_{ml}|^2 \frac
{\re^{-\beta E_{l}} - \re^{-\beta E_{m}}}
{\beta(E_{m}-E_{l})}[\beta|E_{m}-E_{l}|]^{(2n+1)q} \right\}^{1/q}.
\label{estH1}
\end{eqnarray}

One of the possible choices of $p$ and $q$ here, namely even
integer $q=2k$ (hence, $p= 2k/(2k-1)$) leads to the set of
generalized Ginibre inequalities ($k=1,2,3,\dots$):
\begin{equation}
F_{2n}(J;J)\leqslant  \frac{1}{2}F_{2n+1}(J;J)\leqslant  F_{2n}(J;J)+
\frac{1}{2}(J;J)^{(2k-1)/2k}[F_{2k(2n+1)}(J;J)]^{1/2k}.
\label{Ggen}
\end{equation}

At $n=0$ the above set reduces to a symmetric version of the
inequalities obtained by Plechko~\cite{P}:
\begin{equation}
(J;J)\leqslant  \frac{1}{2}\langle JJ^{+} +J^{+}J\rangle \leqslant  (J;J)+
\frac{1}{2}(J;J)^{(2k-1)/2k}\beta (R_k;R_k)^{1/2k}, \quad
(k=1,2,3,\dots) . \label{Plechko}
\end{equation}
Hence, in the particular case of $k=1$ one obtains the Ginibre
inequality (\ref{G}).

\subsection{Generalization of the Bogoliubov~(Jr.)-Plechko-Repnikov
inequalities}

If in (\ref{estH1}) one chooses odd $q=2k+1$, hence,
$p=(2k+1)/2k$, then for the sum in the right-hand side one can use
the upper bound
\begin{eqnarray}
\lefteqn{Z^{-1}\sum_{ml}|J_{ml}|^2 \frac {\re^{-\beta E_{l}} - \re^{-\beta
E_{m}}} {\beta(E_{m}-E_{l})}[\beta|E_{m}-E_{l})|]^{(2n+1)(2k+1)}}
\nonumber \\&& \leqslant  Z^{-1}\sum_{ml}|J_{ml}|^2 |\re^{-\beta E_{l}} +
\re^{-\beta E_{m}}| [\beta|E_{m}-E_{l})|]^{2(2nk+n+k)}
=F_{2(2nk+n+k)+1}(J;J). \label{estJr}
\end{eqnarray}
Thus one obtains the set of inequalities ($k=1,2,3,\dots$)
\begin{equation}
\frac{1}{2}F_{2n+1}(J;J)\leqslant
F_{2n}(J;J)+\frac{1}{2}(J;J)^{2k/(2k+1)}[F_{2(2nk+n+k)+1}(J;J)]^{1/(2k+1)}.
\label{genJr}
\end{equation}

At $n=0$ these reduce to a symmetric version of the set of
inequalities obtained by \linebreak Bogoliubov~Jr., Plechko and Repnikov
\cite{BPR}:
\begin{equation}
\frac{1}{2}\langle JJ^+ + J^+J\rangle \leqslant
(J;J)+\frac{1}{2}(J;J)^{2k/(2k+1)}\{\beta^{2k}\langle R_k R_k^{+}
+ R_k^{+} R_k\rangle\}^{1/(2k+1)}. \label{BPR}
\end{equation}

The symmetric version of the Bogoliubov's (Jr.) inequality
(\ref{Jr}) follows from here in the particular case of $k=1$.

\subsection{Alternative sets of inequalities}

By applying the H\"{o}lder inequality (\ref{Holder}) to the second
term in the right-hand side of (\ref{BHgen}) under the
substitution
\begin{eqnarray}
&&|x_{ml}| = \left\{|J_{ml}|^2 \frac {\re^{-\beta E_{l}} - \re^{-\beta
E_{m}}}{\beta(E_{m}-E_{l})}[\beta(E_{m}-E_{l})]^{2n}\right\}^{1/p}, \nonumber \\
&&|y_{ml}| = \left\{|J_{ml}|^2 |\re^{-\beta E_{l}} - \re^{-\beta
E_{m}}| [\beta|E_{m}-E_{l}|]^{2n+q/p}\right\}^{1/q}, \label{H2}
\end{eqnarray}
instead of (\ref{H1}), one can in parallel derive two new sets of
inequalities. Thus we obtain first
\begin{eqnarray}
\lefteqn{(2Z)^{-1}\sum_{ml}|J_{ml}|^{2}|\re^{-\beta E_{l}} - \re^{-\beta
E_{m}}|[\beta(E_{m}-E_{l})]^{2n}}\nonumber \\ &&\leqslant
\frac{1}{2}[F_{2n}(J;J)]^{1/p}\left\{Z^{-1}\sum_{ml}|J_{ml}|^2
\frac {\re^{-\beta E_{l}} - \re^{-\beta E_{m}}}
{\beta(E_{m}-E_{l})}[\beta|E_{m}-E_{l}|]^{2n+1/(p-1)}\right\}^{(p-1)/p}.\qquad
\label{estH2}
\end{eqnarray}

Now there are two choices of $p$, one of which yields $1/(p-1)$
odd integer, and the other -- even integer. In the first case we
set $1/(p-1)= 2k-1$, $k=1,2,3,\dots$, which implies $p=2k/(2k-1)$,
$q=2k$. Then, from (\ref{estH2}) and (\ref{BHgen}) the following
set of inequalities follows
\begin{equation}
\frac{1}{2}F_{2n+1}(J;J)\leqslant
F_{2n}(J;J)+\frac{1}{2}[F_{2n}(J;J)]^{(2k-1)/2k}[F_{2(n+k)}(J;J)]^{1/2k},\qquad
(k=1,2,3,\dots).\,  \label{genT1}
\end{equation}

In the second case we set $1/(p-1)= 2k$, $k=1,2,3,\dots$, which
implies $p=(2k+1)/2k$, $q=2k+1$. Then we can use the upper bound
\begin{eqnarray}
\lefteqn{ Z^{-1}\sum_{ml}|J_{ml}|^2 |\re^{-\beta E_{l}} - \re^{-\beta E_{m}}|
[\beta|E_{m}-E_{l})|]^{2(n+k)}} \nonumber \\&& \leqslant
Z^{-1}\sum_{ml}|J_{ml}|^2 |\re^{-\beta E_{l}} + \re^{-\beta E_{m}}|
[\beta|E_{m}-E_{l})|]^{2(n+k)} =F_{2(n+k)+1}(J;J), \label{estT2}
\end{eqnarray}
to obtain another set of inequalities
\begin{equation}
\frac{1}{2}F_{2n+1}(J;J)\leqslant
F_{2n}(J;J)+\frac{1}{2}[F_{2n}(J;J)]^{2k/(2k+1)}[F_{2(n+k)+1}(J;J)]^{1/(2k+1)},\qquad
(k=1,2,3,\dots). \label{genT2}
\end{equation}

Note that (\ref{genT1}) is different from (\ref{Ggen}) but at
$n=0$ it reduces to the Plechko inequalities~(\ref{Plechko}).
Therefore, this set of inequalities can also be considered as a
generalization of the Ginibre inequality~(\ref{G}).

Similarly, for general $n$, equation (\ref{genT2}) differs from
(\ref{genJr}) but reduces to the Bogoliubov~(Jr.)-Plechko-Repnikov
inequalities (\ref{BPR}) at $n=0$. Hence, the set of inequalities
(\ref{genT2}) is also a generalization of the Bogoliubov (Jr.)
inequality (\ref{Jr}).

A comment is in order here. Due to the property
$R_{n+k}(J)=R_{k}(R_{n}(J))$, in terms of the operators $B_n
\equiv R_{n}(J)$ one has
\begin{eqnarray}
&&F_{2n}(J;J)= \beta^{2n}(B_n;B_n), \quad F_{2n+1}(J;J)=
\beta^{2n}\langle B_n B_n^+ + B_n^+ B_n\rangle, \nonumber \\&&
F_{2(n+k)}(J;J)= \beta^{2(n+k)}(R_k(B_n);R_k(B_n)),\nonumber \\&&
F_{2(n+k)+1}(J;J)= \beta^{2(n+k)}\langle R_k(B_n)R_k^\dagger(B_n)+
R_k^\dagger(B_n)R_k(B_n)\rangle. \label{PartT}
\end{eqnarray}
Therefore, inequalities (\ref{BHgen}) take exactly the form of the
inequality of Brooks Harris (\ref{Harris}) with $A$ replaced by
$B_n$, inequalities (\ref{genT1}), respectively (\ref{genT2}),
take exactly the form of a symmetric version of the Plechko
inequalities~(\ref{Plechko}), respectively, the
Bogoliubov~(Jr.)-Plechko-Repnikov inequalities~(\ref{BPR}), with
$J$ replaced by $B_n$.

Notably, under the above substitution, the generalized Ginibre
inequalities (\ref{Ggen}) and the generalized
Bogoliubov~(Jr.)-Plechko-Repnikov inequalities (\ref{genJr}) do
not reduce to any of the known types of inequalities, except in
the particular case of $n=0$.

\subsection{General features and comparison of upper bounds}

Obviously, the main inequalities have been derived under different
approximations. The most direct is the derivation of generalized
Harris inequalities - it is based upon the single elementary upper
bound (\ref{elemBH}). Next, the generalized Ginibre inequalities
are derived by first using the elementary upper bound
(\ref{elemG}), followed by the application of the H\"{o}lder
inequality (\ref{Holder}) under a special choice of the
parameters: $p=2k/(2k-1)$ and $q=2k$. An alternative choice of
these parameters, $p=(2k+1)/2k$ and $q=2k+1$, requires the use of
an additional, rather crude upper bound (\ref{ub}), in order to
derive the Bogoliubov~(Jr.)-Plechko-Repnikov inequalities
(\ref{genJr}).

The characteristic feature of our generalized inequalities is
that, under sufficiently mild conditions on the Hamiltonian
${\mathcal H}_N$ of the $N$-particle system and the bounded
intensive observable $J_N$, they provide the upper bounds on the
non-negative difference
\begin{equation}
0 \leqslant  \Delta_n(J_N) \equiv \frac{1}{2}\langle R^+_n(J_N)R_n(J_N)
+ R_n(J_N)R^+_n(J_N)\rangle_{{\mathcal H}_N} -
(R_n(J_N);R_n(J_N))_{{\mathcal H}_N} \label{difRn}
\end{equation}
of equal form and equal order of magnitude (with respect to $N$)
for all finite $n=0,1,2,\dots $, i.e., for any finite set of
observables $J_N$, $R_1(J_N) = [{\mathcal H}_N, J_N]$, $R_2(J_N) =
[{\mathcal H}_N, R_1(J_N)], \dots$. Note that the left-hand side
inequality in (\ref{difRn}) is a generalization of the convexity
property of the Bogoliubov-Duhamel inner product (\ref{conv}).

The required conditions are ($n=1,2,3,\dots $):
\begin{eqnarray}
|\langle J_N \rangle_{{\mathcal H}_N}| &\leqslant&  O(1), \nonumber \\
F_{2n}(J_N;J_N) &=& \beta^{2n-1} \langle [R^+_n(J_N),R_{n-1}(J_N)]\rangle_{{\mathcal H}_N}
\leqslant  O(N^{-1}), \nonumber \\
F_{2n+1}(J_N;J_N) &=& \beta^{2n}\langle R^+_n(J_N)R_n(J_N) +
R_n(J_N)R^+_n(J_N)\rangle_{{\mathcal H}_N} \leqslant  O(1).
\label{gencond}
\end{eqnarray}
They are generally satisfied for extensive Hamiltonians ${\mathcal
H}_N$ with bounded interaction and bounded intensive observables
$J_N$, which are arithmetic averages over the particles of the
system of some local observables. An example will be given in the
end of this section.

The above conditions may also hold in some cases when ${\mathcal
H}_N$ and $R_n(J_N)$, $n=0,1,2,\dots$, contain unbounded
operators. As an example in section~\ref{Dicke} we will consider
the Dicke model of superradiance~\cite{Dicke}, for which the
proper unbounded counterpart of $J_N$ will be found.

Under conditions (\ref{gencond}), our generalized inequalities yield the following upper
bounds for all finite $n=0,1,2,\dots $:
\begin{enumerate}
\item%
 Generalized Harris inequality (\ref{BHgen})
\begin{equation}
\Delta_n(J_N) \leqslant  O(N^{-1}).
\label{Harrisbound}
\end{equation}
\item%
 Generalized Plechko inequalities (\ref{Ggen})
\begin{equation}
\Delta_n(J_N) \leqslant  (J_N;J_N)^{(2k-1)/2k}O(N^{-1/2k}),\qquad
(k=1,2,3,\dots), \label{Plechkobound}
\end{equation}
which at $k=1$ reduce to the Ginibre inequality
\begin{equation}
\Delta_n(J_N) \leqslant  (J_N;J_N)^{1/2}O(N^{-1/2}).
\label{Ginibrebound}
\end{equation}
\item%
Generalized Bogoliubov~(Jr.)-Plechko-Repnikov inequalities
(\ref{genJr})
\begin{equation}
\Delta_n(J_N) \leqslant  (J_N;J_N)^{2k/(2k+1)}O(1),\qquad (k=1,2,3,\dots),
\label{BPRbound}
\end{equation}
which at $k=1$ reduce to the Bogoliubov (Jr.) inequality
\begin{equation}
\Delta_n(J_N) \leqslant  (J_N;J_N)^{2/3}O(1).
\label{Jrbound}
\end{equation}
\end{enumerate}

Due to the relationship between the Bogoliubov-Duhamel inner
product and the susceptibility of the system with respect to the
external field conjugate to the observable $J_N$, see
(\ref{second}), we can compare the different upper bounds in the
region of parameters in which the susceptibility is bounded. We
see that the generalized Harris, Plechko and Ginibre inequalities
yield upper bounds of the order $O(N^{-1})$, while the generalized
Bogoliubov~(Jr.)-Plechko-Repnikov inequalities yield upper bounds
of the order $O(N^{-2k/(2k+1)})$, $k=1,2,3,\dots$.

Finally, to illustrate the validity of conditions (\ref{gencond})
and the explicit form of the first generalized observables
$R_n(J_N)$, we consider a simple model with separable attraction
built upon bounded operators. Let the model system contain $N$
spins and have the Hamiltonian (normalized by $k_{\rm B} T$)
\begin{equation}
\beta {\mathcal H}_N= -Ng_x \left(J^x_N\right)^2 -Ng_y
\left(J^y_N\right)^2 - N \mathbf{h\cdot J}_N\,,
\label{BoudedH}
\end{equation}
where $g_x,\; g_y >0$ are dimensionless coupling constants,
$\mathbf{h}= \{h_x,h_y,h_z\}$ is the vector of the external magnetic
field, $\mathbf{J}_N =\{J_N^x,J_N^y,J_N^z\}$ is the vector operator
of the average spin with uniformly bounded in $N$ components $J_N^\alpha$,
\begin{equation}
J_N^\alpha = \frac{1}{N}\sum_{i =1}^N \sigma_i^\alpha\,, \qquad \alpha
=x,y,z, \label{defJ}
\end{equation}
where $\sigma^\alpha$ are the standard Pauli matrices.

By using the commutation relations for the spin operators, we
obtain the observables
\begin{equation}
R_1(J_N^x) \equiv [\beta {\mathcal H}_N, J_N^x] =
2\mathrm{i}[g_y(J_N^y J_N^z + J_N^z J_N^y) +h_y
J_N^z - h_z J_N^y], \label{R1J}
\end{equation}
and
\begin{eqnarray}
R_2(J_N^x) \equiv [\beta {\mathcal H}_N, R_1(J_N^x)]& =&
4g_y(g_y -g_x)(J_N^y J_N^y J_N^x +
2J_N^y J_N^x J_N^y + J_N^x J_N^y J_N^y) \nonumber \\
&&{}+ 4g_x g_y(J_N^z J_N^z J_N^x + 2J_N^z J_N^x J_N^z +
J_N^x J_N^z J_N^z) \nonumber
\\ &&{}+ 4(g_y -g_x)h_y (J_N^x J_N^y + J_N^y J_N^x) -4g_x h_z
(J_N^x J_N^z + J_N^z J_N^x)
\nonumber \\&&{}+ 4(h_y^2 +h_z^2)J_N^x -4h_x h_y J_N^y - 4h_x h_z J_N^z\,,
\label{R2J}
\end{eqnarray}
which, by definition, have zero average values. Next,
\begin{eqnarray}
F_2(J_N^x;J_N^x) &\equiv& (R_1(J_N^x); R_1(J_N^x)) = \langle [R_1^+(J_N^x),
J_N^x]\rangle \nonumber \\&=& \frac{4}{N}\left\{2g_y [\langle
(J^y_{N})^2\rangle - \langle (J^z_{N})^2\rangle ] + h_y \langle J_N^y\rangle
+ h_z \langle J_N^z\rangle \right\}, \label{F2J}
\end{eqnarray}
and
\begin{eqnarray}
F_3(J_N^x;J_N^x) &\equiv &  \langle [R_1^+(J_N^x) R_1(J_N^x) +
R_1(J_N^x)R_1^+(J_N^x)]\rangle \nonumber \\&=& \langle [g_y(J^y J^z +
J^z_{N} J^y_{N}) +h_y J^z_{N} - h_z J^y_{N}]^2\rangle. \label{F3J}
\end{eqnarray}

Rather lengthy but straightforward calculations show that
\begin{equation}
|F_4(J_N^x;J_N^x)| = |\langle [R_2^+ (J_N^x), R_1(J_N^x)]\rangle| \leqslant
O(N^{-1}). \label{F4J}
\end{equation}

One can readily extend the above results and show that for all
finite $n$ the following inequalities hold: $|\langle
R_n(J_N^x)\rangle | \leqslant  ||R_n(J_N^x)||\leqslant  O(1)$,
$|F_{2n}(J_N^x;J_N^x)| \leqslant  O(N^{-1})$, and
$|F_{2n+1}(J_N^x;J_N^x)| \leqslant  O(1)$. Thus, conditions
(\ref{gencond}) and, hence, the upper bounds (1)--(3) are
fulfilled for the considered model.

\section{Systems of matter interacting with Boson fields}

\subsection{Models and their treatment by the AHM}

Here are two examples of models in solid state physics, which
belong to this class: (a) The Dicke model of superradiance
\cite{Dicke}, solved exactly in~\cite{HL73a,HL73b}, and by the AHM
in~\cite{BZT75}. The model has been generalized to include
interactions with both electromagnetic field and phonons
\cite{KZ77,KZ78}, and to the case of infinitely many modes of the
electromagnetic field~\cite{Z84}. A recent review  of the
thermodynamic properties of the  original Dicke model and its
generalizations  is given in~\cite{TBZ09}. (b) The Mattis-Langer
model of structural instability~\cite{ML70}, solved exactly by the
AHM in~\cite{BTZ75}. A class of models, including as a special
case the Dicke model, has been considered in the framework of the
AHM by Bogoliubov (Jr.) and Plechko~\cite{BP75}. The
one-dimensional case with countably infinite set of phonon modes
has been solved by means of theta-function integration in
\cite{BP84}.

The model Hamiltonian is defined on the tensor product of two
Hilbert spaces, one for the subsystem describing matter (e.g.,
electrons in a solid, spins on a lattice), and the other for the
Boson field (lattice vibrations, electromagnetic field). In the
second quantization representation, the creation,
$b_{\mathrm{s}}^{\dagger}$, and annihilation, $b_{\mathrm{s}}$,
operators of the Boson field modes (labeled by the subscript~$s$)
satisfy the canonical commutation relations
\begin{equation}
[b_{\mathrm{s}}, b_l^{\dagger}]=\delta_{s,l},\qquad [b_{\mathrm{s}}, b_l] =
[b_{\mathrm{s}}^{\dagger}, b_l^{\dagger}] =0 , \label{1c137}
\end{equation}
for all the allowed $s$ and $l$. Since the Boson operators are
unbounded, it is not possible to obtain easy bounds on their
average values in terms of Hilbert-space norm.

Another characteristic feature of this class of models is that
exact solvability by the AHM is possible when the interaction with
only a finite (or growing slower than the volume $V$, as $V
\rightarrow \infty$) number of Boson modes is taken into account.
The typical model Hamiltonian has the form
\begin{equation}
{\mathcal H}_{\Lambda} = {\mathcal T}_{\Lambda}+\sum_{s=1}^{n}
\omega_{\mathrm{s}} \; b_{\mathrm{s}}^{\dagger} b_{\mathrm{s}}+V^{1/2} \sum_{s=1}^{n}\lambda_{\mathrm{s}} (b_{\mathrm{s}}
A_{s,\Lambda}^{\dagger}+ b_{\mathrm{s}}^{\dagger}A_{s,\Lambda}).
\label{1c138}
\end{equation}
Here the operators ${\mathcal T}_{\Lambda} = {\mathcal
T}_{\Lambda}^{\dagger}$ and $A_{s,\Lambda}$, $s=1,\dots , n$,
refer to the matter subsystem and satisfy the general conditions
(\ref{2.2}) and (\ref{2.4}). The second term in the right-hand
side of (\ref{1c138}) describes a finite number of free Boson
modes, $s=1,\dots , n$, in the space domain $\Lambda $ of volume
$V$. For the sake of simplicity, the energies $\omega_{\mathrm{s}}
>0$ and the interaction constants $\lambda_{\mathrm{s}} \in R$ are
taken to be independent of the volume $|\Lambda|$. The
mathematical definition of the above Hamiltonian is given
in~\cite{ZBT76}.

The corresponding approximating Hamiltonian depends on a set of
complex numbers $\eta =(\eta_1,\dots , \eta_n) \in {\mathbf C}^n$
and has the form, see~\cite{BZT75,BBZKT84},
\begin{equation}
{\mathcal H}_{\Lambda}^{(0)}(\eta)=\sum_{s=1}^{n} \omega_{\mathrm{s}} \;
\tilde{b}_{\mathrm{s}}^{\dagger}\tilde{b}_{\mathrm{s}} + {\mathcal T}_{\Lambda}
-V\sum_{s=1}^{n} (\lambda_{\mathrm{s}}^2 /\omega_{\mathrm{s}})(\eta_{\mathrm{s}}
A_{s,\Lambda}^{\dagger}+ \eta_{\mathrm{s}}^{\star}  A_{s,\Lambda} -
\eta_{\mathrm{s}}^{\star} \eta_{\mathrm{s}}), \label{1c151}
\end{equation}
where
\begin{equation}
\tilde{b}_{\mathrm{s}}^{\dagger}= b_{\mathrm{s}}^{\dagger}
+V^{1/2}\frac{\lambda_{\mathrm{s}}}{\omega_{\mathrm{s}}} \eta_{\mathrm{s}}^{\star}\,, \qquad
\tilde{b}_{\mathrm{s}} =b_{\mathrm{s}} +V^{1/2}\frac{\lambda_{\mathrm{s}}}{\omega_{\mathrm{s}}}\eta_{\mathrm{s}}\,,
\label{shiftb}
\end{equation}
are the creation/annihilation operators for a subsystem of free
shifted bosons. The application of the Bogoliubov inequalities to
the difference of the free energy densities for the model and
approximating Hamiltonians yields
\begin{equation}
f[{\mathcal H}_{\Lambda}^{(0)}(\eta)] - f[{\mathcal H}_{\Lambda}]
\geqslant 0, \label{minf}
\end{equation}
for all $\eta \in {\mathbf C}^n$. Therefore, the best
approximation is reached at $\eta = \bar{\eta}_{\Lambda}$, where
$\bar{\eta}_{\Lambda}$ satisfies the absolute minimum condition
\begin{equation}
f[{\mathcal H}_{\Lambda}^{(0)}(\bar{\eta}_{\Lambda})]= \min_{\eta}
f[{\mathcal H}_{\Lambda}^{(0)}(\eta)]. \label{abscond}
\end{equation}

Note that the free energy density of the free bosons,
\begin{equation}
f\left[\sum_{s=1}^{n} \omega_{\mathrm{s}} \; b_{\mathrm{s}}^{\dagger} b_{\mathrm{s}}\right]=
\frac{1}{\beta V}\sum_{s=1}^{n}\ln \left(1-{\mathrm e}^{-\beta
\omega_{\mathrm{s}}}\right) , \label{bosonf}
\end{equation}
is independent of the parameters $\eta =(\eta_1,\dots , \eta_n)
\in {\mathbf C}^n$ and vanishes in the thermodynamic limit as
$O(V^{-1})$.

The proof of the thermodynamic equivalence of the free energy
densities $f[{\mathcal H}_{\Lambda}^{(0)}(\bar{\eta}_{\Lambda})]$
and $f[{\mathcal H}_{\Lambda}]$ goes again through the
introduction of sources of the boson fields,
\begin{eqnarray}
{\mathcal H}_{\Lambda}(\nu) &=& {\mathcal H}_{\Lambda} - V^{1/2}
\sum_{s=1}^{n}(\nu_{\mathrm{s}}^{\star} b_{\mathrm{s}} + \nu_{\mathrm{s}} b_{\mathrm{s}}^{\dagger}), \nonumber
\\ {\mathcal H}_{\Lambda}^{(0)}(\eta, \nu)&=&{\mathcal H}_{\Lambda}^{(0)}(\eta)
- V^{1/2} \sum_{s=1}^{n}(\nu_{\mathrm{s}}^{\star} b_{\mathrm{s}} + \nu_{\mathrm{s}} b_{\mathrm{s}}^{\dagger}),
\label{Hnu}
\end{eqnarray}
where $\nu =(\nu_1,\dots , \nu_n) \in {\mathbf C}^n$.

Now, from the Bogoliubov inequalities and a subsequent use of an
elementary upper bound, one obtains
\begin{eqnarray}
0&\leqslant & \min_{\eta}f[{\mathcal H}_{\Lambda}^{(0)}(\eta,\nu)]-
f[{\mathcal H}_{\Lambda}(\nu)] \leqslant
-V^{-1/2}\sum_{s=1}^{n}\lambda_{\mathrm{s}} \langle \delta b_{\mathrm{s}} \delta
A_{s,\Lambda} + \delta b_{\mathrm{s}}^{\dagger} \delta
A_{s,\Lambda}\rangle_{{\mathcal H}_{\Lambda}(\nu)} \nonumber
\\&\leqslant & V^{-1/2}\sum_{s=1}^{n}V^{-\gamma}
\frac{\lambda_{\mathrm{s}}^2}{\omega_{\mathrm{s}}} \langle \delta
A_{s,\Lambda}^{\dagger} \delta A_{s,\Lambda}\rangle_{{\mathcal
H}_{\Lambda}(\nu)} +V^{-1/2}\sum_{s=1}^{n} V^{\gamma} \omega_{\mathrm{s}}
\langle \delta b_{\mathrm{s}}^{\dagger} \delta b_{\mathrm{s}} \rangle_{{\mathcal
H}_{\Lambda}(\nu)}, \label{upb}
\end{eqnarray}
where
\begin{equation}
\delta A_{s,\Lambda} = A_{s,\Lambda} -\langle
A_{s,\Lambda}\rangle_{{\mathcal H}_{\Lambda}(\nu)}, \qquad \delta
b_{\mathrm{s}} = b_{\mathrm{s}} - \langle b_{\mathrm{s}}\rangle_{{\mathcal H}_{\Lambda}(\nu)},
\end{equation}
and $\gamma \in (0,1)$ is a free parameter.

Due to the boundedness of the operators $A_{s,\Lambda}$,
$A_{s,\Lambda}^\dagger$, see conditions (\ref{2.4}), the first sum
in the right-hand side of the last inequality (\ref{upb}) is
bounded from above by
\begin{equation}
V^{-1/2-\gamma}n M_1^2 \max_{\mathrm{s}} (\lambda_{\mathrm{s}}^2/\omega_{\mathrm{s}}).
\end{equation}

The quadratic fluctuations of the boson fields, $\langle \delta
b_{\mathrm{s}}^{\dagger} \delta b_{\mathrm{s}} \rangle_{{\mathcal H}_{\Lambda}(\nu)}$,
are to be majorized by terms proportional to powers of the
Bogoliubov-Duhamel inner product
\begin{equation}
(\delta b_{\mathrm{s}}; \delta b_{\mathrm{s}})_{{\mathcal H}_{\Lambda}(\nu)}=
-\frac{1}{\beta}\frac{\partial^2 f[{\mathcal H}_{\Lambda}(\nu)]}
{\partial \nu_{\mathrm{s}} \partial \nu_{\mathrm{s}}^\star}\,.
\end{equation}

By using the Ginibre inequality (\ref{G}) and choosing $\gamma =
1/3$ in (\ref{upb}), the subsequent application of the
majorization technique due to Bogoliubov Jr. with the following
bounds on the first derivatives ($s=1,\dots,n$)
\begin{equation}
\left|\frac{\partial f[{\mathcal H}_{\Lambda}(\nu)]} {\partial
\nu_{\mathrm{s}}^{\#}}\right| = V^{-1/2}| \langle b_{\mathrm{s}}^{\#}\rangle_{{\mathcal
H}_{\Lambda}(\nu)} |= \omega_{\mathrm{s}}^{-1}|\lambda_{\mathrm{s}} \langle
A_{s,\Lambda}^{\#}\rangle_{{\mathcal H}_{\Lambda}(\nu)} -
\nu_{\mathrm{s}}^{\#}|\leqslant  \omega_{\mathrm{s}}^{-1}(\lambda_{\mathrm{s}} M_1 + |\nu_{\mathrm{s}}^{\#}|),
\label{bound1}
\end{equation}
made it possible to prove that
\begin{equation}
|\min_{\eta} f[{\mathcal H}_{\Lambda}^{(0)}(\eta)] - f[{\mathcal
H}_{\Lambda}]| \leqslant  \epsilon_V, \label{1c169}
\end{equation}
where $\epsilon_V = O(V^{-1/3}) \rightarrow  0$ as $V \rightarrow
\infty$. Under additional conditions on the double commutators
between the different $\{A_{s,\Lambda}^{\#}: s=1,\dots,n\}$ and on
the commutator of $R_1(A_{s,\Lambda}^{\#})$ with
$A_{s',\Lambda}^{\#}$, the above estimate was improved up to
$\epsilon_V = O(V^{-1/2})$~\cite{BBZKT81}.

\subsection{Application of the generalized inequalities to
the Dicke model}\label{Dicke}

In the remainder, by using the equality~\cite{G,DLS}:
\begin{equation}
\beta (X; [{\mathcal H},B])_{\mathcal H}=\langle
[X^\dagger,B]\rangle_{\mathcal H}\,, \label{rel2}
\end{equation}
we shall derive a variety of explicit relationships between
Bogoliubov-Duhamel inner products and usual thermal averages for
different observables of the basic single-mode Dicke model in the
rotating wave approximation. In the latter case, the Hamiltonian
has the form (\ref{1c138}) with $n=1$,
\begin{equation}
T = \frac{1}{2}\epsilon \sum_{j=1}^N \sigma_j^z\,,\qquad A =
\frac{1}{V}\sum_{j=1}^N \sigma_j^+\,,\qquad A^\dagger =
\frac{1}{V}\sum_{j=1}^N \sigma_j^-\,, \label{TAA}
\end{equation}
where $\sigma_j^\pm = \frac{1}{2}(\sigma_j^x \pm
\mathrm{i}\sigma_j^y)$ and $\sigma_j^z$ are the Pauli matrices.
From the above definitions the following commutation relations
follow
\begin{equation}
[T,A] = \epsilon A, \qquad [A^\dagger ,A]= \frac{2}{\epsilon V^2}T.
\end{equation}

By direct computation we obtain
\begin{equation}
R_1(b) \equiv [H, b] = -(\omega b + V^{1/2}\lambda A), \label{R1b}
\end{equation}
and, since $\langle [H, b]\rangle_{\mathcal H} =0$, we obtain a
well known equality between average values of observables
pertaining to different subsystems (boson field and matter):
\begin{equation}
\langle b\rangle_{\mathcal H} = - V^{1/2}\frac{\lambda}{\omega}
\langle A\rangle_{\mathcal H}\,. \label{rel1b1}
\end{equation}
Therefore, in the case of non-vanishing polarization in the matter
subsystem, $\langle A\rangle_{\mathcal H}\not= 0$, the average
value of the boson annihilation (as well as creation) operator
will behave as $\langle b\rangle_{\mathcal H}=O(V^{1/2})$. Hence,
we deduce that the unbounded counterpart of the observable $J_N$
in conditions (\ref{gencond}) should be the normalized operator
$V^{-1/2}b$. Keeping this in mind, we further calculate
\begin{equation}
R_2(V^{-1/2}b) \equiv [H, R_1(V^{-1/2}b)] = \omega^2 V^{-1/2}b +
\lambda (\omega - \epsilon)A - \frac{2\lambda^2}{\epsilon V^{3/2}}
b T, \label{R2b}
\end{equation}
and
\begin{eqnarray}
F_2(V^{-1/2}b;V^{-1/2}b)&=&\beta \langle
[R_1(V^{-1/2}b)^\dagger,V^{-1/2}b]\rangle_{\mathcal H}
=\frac{\beta \omega}{V}\,,
\label{F2b}\\
F_3(V^{-1/2}b;V^{-1/2}b)&\equiv& \beta^2\langle
R_1^\dagger(V^{-1/2}b) R_1(V^{-1/2}b)+
R_1(V^{-1/2}b)R_1^\dagger(V^{-1/2}b)
\rangle_{\mathcal H} \nonumber \\
& =& (\beta \omega)^2\left[V^{-1}\langle b^\dagger b +b b^\dagger
\rangle_{\mathcal H} + V^{-1/2}\frac{\lambda}{\omega}\langle
b^\dagger A + b A^\dagger\rangle_{\mathcal H} +
\frac{\lambda^2}{\omega^2}\langle A^\dagger A + A A^\dagger
\rangle_{\mathcal H} \right],\nonumber\\ \label{F3b}\\
F_4(V^{-1/2}b;V^{-1/2}b)&=&\beta^3\langle
[R_2^\dagger(V^{-1/2}b),R_1(V^{-1/2}b)]\rangle_{\mathcal H} \nonumber \\
&=& \frac{(\beta\omega)^3}{V}\left[1 +\frac{2\lambda^2}{\epsilon
\omega^2 V} \left(\frac{\epsilon}{\omega} -2\right)\langle
T\rangle_{\mathcal H} + \frac{2\lambda^3}{\omega^3 V^{1/2}}\langle
b^\dagger A\rangle_{\mathcal H} \right]. \label{F4b}
\end{eqnarray}

The right-hand side of these expressions can be evaluated using
some relationships between average values of different observables
which follow from (\ref{rel2}). Thus, by setting $B=b$ and $X=b$,
we obtain,
\begin{equation}
(b;b)_{\mathcal H}= \frac{1}{\beta \omega} -
V^{1/2}\frac{\lambda}{\omega} (b;A)_{\mathcal H} \,. \label{rel1b2}
\end{equation}
Next, from $B=b$ and $X=A$ it follows that
\begin{equation}
-(A;b)_{\mathcal H}= V^{1/2}\frac{\lambda}{\omega} (A;A)_{\mathcal
H}\,. \label{rel1b3}
\end{equation}
Under the alternative choice $B=b^\dagger$ and $X=b^\dagger$ in
(\ref{rel2}), we derive
\begin{equation}
(b^\dagger;b^\dagger)_{\mathcal H}= \frac{1}{\beta \omega} -
V^{1/2}\frac{\lambda}{\omega}(b^\dagger;A^\dagger)_{\mathcal H}\,,
\label{rel1b2d}
\end{equation}
which, due to the conjugate symmetry
$(A;B)=(B^\dagger;A^\dagger)$, is equivalent to
\begin{equation}
(b;b)_{\mathcal H}= \frac{1}{\beta \omega} -
V^{1/2}\frac{\lambda}{\omega} (A;b)_{\mathcal H}\, . \label{1b2s}
\end{equation}
By comparing this equality with (\ref{rel1b2}) we conclude that
$(b;A)_{\mathcal H}= (A;b)_{\mathcal H}$. Then, taking into
account (\ref{rel1b3}) we derive the important relation
\begin{equation}
(b;b)_{\mathcal H}= \frac{1}{\beta \omega} +
V\frac{\lambda^2}{\omega^2}(A;A)_{\mathcal H}\,,
 \label{rel1b2s}
\end{equation}
which was obtained in~\cite{TBZ09} using gage invariance
arguments.

Proceeding further with the evaluation of (\ref{F3b}) and
(\ref{F4b}), we note that since the right-hand side of (\ref{F4b})
is real, one must have
\begin{equation}
\langle b^\dagger A\rangle_{\mathcal H}= \langle b A^\dagger
\rangle_{\mathcal H}\,. \label{bAAb}
\end{equation}
Therefore, the application of the Schwarz inequality and the
relation (\ref{rel1b2s}) yield the estimate
\begin{equation}
|\langle V^{-1/2}b^\dagger A\rangle_{\mathcal H}|= \langle
V^{-1/2}b A^\dagger \rangle_{\mathcal H}| \leqslant  \langle
V^{-1}b^\dagger b\rangle^{1/2}_{\mathcal H}| \langle A A^\dagger
\rangle^{1/2}_{\mathcal H}| \leqslant  O(1). \label{Schwarz}
\end{equation}

Finally, the application of the Harris inequality (\ref{Harris})
in terms of  $b$ gives
\begin{equation}
(b;b)_{\mathcal H} \leqslant  \langle b^\dagger b \rangle_{\mathcal H} +
\frac{1}{2} \leqslant  (b;b)_{\mathcal H} +\frac{1}{12}\beta \omega \,,
 \label{bb}
\end{equation}
which, in view of (\ref{rel1b2s}), implies $\langle b^\dagger b
\rangle_{\mathcal H} = O(V)$, provided that $(A;A)_{\mathcal H} =
O(1)$. Taking into account $||A|| = O(1)$ and $||T|| = O(V)$, we
conclude that $F_2(V^{-1/2}b;V^{-1/2}b)$ and
$F_4(V^{-1/2}b;V^{-1/2}b)$ are bounded from above by $O(V^{-1})$,
while $|F_3(V^{-1/2}b;V^{-1/2}b)| \leqslant  O(1)$.

Thus, we have proved that $V^{-1/2}b$ and
$F_n(V^{-1/2}b;V^{-1/2}b)$ for $n=2,3,4$ satisfy
conditions~(\ref{gencond}). Therefore, bounds of the form
(\ref{Harrisbound}), (\ref{Ginibrebound}), and (\ref{BPRbound})
hold true at $n=1$ even for the unbounded operator $J_V =
V^{-1/2}b$.

\section{Discussion}
\looseness=-1 The AHM makes it possible to rigorously obtain the
exact thermodynamic properties of diverse classes of model systems
in quantum statistical mechanics. Each  class of models  requires
a certain structure of the interaction Hamiltonian and special
properties of the operators in terms of which its structure is
defined. Initially, the mathematical technique for derivation of
bounds which prove the thermodynamic equivalence of the
approximating and original Hamiltonians has been developed for
systems with interaction Hamiltonians constructed with bounded
operators (see section~2). Later, it was extended to the case of
unbounded Bose operators (see section~4). This technique
essentially exploits the possibility  of estimating the
correlation functions in~(\ref{kf}) or~(\ref{upb}) from above by
expressions containing the Bogoliubov-Duhamel inner product in
combination with  average values of certain operators. As a rule,
the latter are estimated by norm, which requires conditions
(\ref{2.2})--(\ref{2.93}). It is well known that this procedure is
not unique~\cite{BBZKT81,BDT}. The distinction comes from the
different inequalities used. At first, it was Bogoliubov's Jr.
inequality that was used in the case of bounded operators. Latter
it was realized  that Ginibre's inequality (\ref{G}) is more
convenient to be used, especially in the case of interactions
involving Bose operators~\cite{BBZKT81}, see also section~4.
Moreover, it made it possible to derive a better estimate, namely
$O(|\Lambda|^{-1/2})$, in both cases of bounded and unbounded
operators~\cite{BBZKT84,BBZKT81,BDT}.  However, in the latter
case, two additional sufficient conditions have been imposed.

The interest that stimulates the search and use of different
inequalities was prompted  by the wish to improve the upper bounds
on the difference of the model and approximating free energy
densities and/or to enlarge  the  class of model systems
rigorously solved by the AHM  \cite {P,BPR}. In section~3, we have
derived generalizations of  the inequalities introduced in \cite
{P} and~\cite{BPR}. The novel point is that a set of new
functionals $F_{k}(J;J)$, $k=0,1,2,\dots $, were defined by their
spectral representation (\ref{BT2010}). This made it possible to
obtain different upper bounds in a unified fashion.

Beyond the use in the majorization step of the AHM, our new upper bounds (\ref{Harrisbound})--(\ref{Jrbound})
on the difference between the Gibbs average values of a class of
observables of a many-particle system and the corresponding Bogoliubov-Duhamel inner products may
find wider applications.
Under sufficiently mild conditions, these upper bounds have the same form and order of magnitude
with respect to the number of particles (or volume)
for all the quantities derived by a finite number of
commutations $R_{n}(J)$ of an original intensive observable $J$ with the
Hamiltonian of the system ${\mathcal H}$.

In addition, we have obtained important relationship between average values of the different
observables in the framework of two types of exactly
solved by the AHM model systems: one with bounded separable attraction -- the infinitely coordinated anisotropic
Heisenberg model, and the other, the Dicke superradiance model, describing interaction of a boson field with
a subsystem of matter.

\section*{Acknowledgements}

N.S.T. was supported  by  the National Science Foundation of Bulgaria under grant TK--X--1712/2007.

\newpage
\ukrainianpart

\title{Узагальнені нерівності для внутрішнього добутку Боголюбова-Дюамеля із застосуваннями в методі апроксимуючого гамільтоніану%
}

\author[Дж. Бранков, Н.С. Тончев]{Дж. Г. Бранков\refaddr{label1,label2},
        Н.С. Тончев\refaddr{label3}}

\addresses{
\addr{label1} Лабораторія теоретичної фізики ім. Боголюбова, Об'єднаний інститут ядерних досліджень, \\141980 Дубна, Російська Федерація
\addr{label2} Інститут механіки, Болгарська академія наук, Софія, Болгарія
\addr{label3} Інститут фізики твердого тіла, Болгарська академія наук, Софія, Болгарія
}

\date{отримано 29 листопада 2010}

\makeukrtitle

\begin{abstract}
\tolerance=3000%
Отримано нескінченні набори нерівностей, які узагальнюють всі відомі нерівності, що можуть бути використані на етапі
мажорування методу апроксимуючого гамільтоніану
Вони забезпечують верхні границі на різницю між квадратичними флуктуаціями інтенсивних спостережуваних  $N$-частинкової системи і відповідного внутрішнього добутку Боголюбова-Дюамеля. Новою рисою є те, що при достатньо м'яких умовах верхні границі мають однакову форму і порядок величини по відношенню до $N$ для всіх величин, отриманих шляхом скінченного числа перестановок початкової інтенсивої спостережуваної з гамільтоніаном. Результати ілюструються на двох типах точно розв'язуваних моделей: однієї з обмеженим сепарабельним притяганням та іншої, що містить взаємодію бозонного поля з матерією.
\keywords кореляційні функції, внутрішній добуток Боголюбова-Дюамеля,  нерівності статистичної механіки, метод апроксимуючого гамільтоніану, точно розв'язувані моделі
\end{abstract}

\end{document}